\documentclass{ieeeaccess}
\usepackage{cite}
\usepackage{amsmath,amssymb,amsfonts}

\usepackage{amsmath,amssymb}   
\usepackage{algorithm}         
\usepackage{algpseudocode}     

\usepackage{graphicx}
\usepackage{textcomp}
\usepackage{cite}
\usepackage{hyperref}
\usepackage{newtxtext,newtxmath} 
\usepackage[nameinlink,capitalize]{cleveref}
\usepackage[caption=false,font=normalsize,labelfont=sf,textfont=sf]{subfig}
\usepackage{graphicx}
\usepackage{newtxtext,newtxmath} 
\usepackage{bm}
\makeatletter
\AtBeginDocument{\DeclareMathVersion{bold}
\SetSymbolFont{operators}{bold}{T1}{times}{b}{n}
\SetSymbolFont{NewLetters}{bold}{T1}{times}{b}{it}
\SetMathAlphabet{\mathrm}{bold}{T1}{times}{b}{n}
\SetMathAlphabet{\mathit}{bold}{T1}{times}{b}{it}
\SetMathAlphabet{\mathbf}{bold}{T1}{times}{b}{n}
\SetMathAlphabet{\mathtt}{bold}{OT1}{pcr}{b}{n}
\SetSymbolFont{symbols}{bold}{OMS}{cmsy}{b}{n}
\renewcommand\boldmath{\@nomath\boldmath\mathversion{bold}}}
\makeatother

\def\BibTeX{{\rm B\kern-.05em{\sc i\kern-.025em b}\kern-.08em
    T\kern-.1667em\lower.7ex\hbox{E}\kern-.125emX}}

\begin{document}
\history{Date of publication xxxx 00, 0000, date of current version xxxx 00, 0000.}
\doi{10.1109/ACCESS.2024.0429000}

\title{A Joint Delay-Energy-Security Aware Framework for Intelligent Task Scheduling in Satellite-Terrestrial Edge Computing Network}
\author{\uppercase{TING YOU}\authorrefmark{1}\IEEEmembership{Student Member, IEEE},
\uppercase{YUHAO ZHENG}\IEEEmembership{Graduate Student Member, IEEE}\authorrefmark{2}, \uppercase{KEJIA PENG} \authorrefmark{3, 4}, AND \uppercase{CHANG LIU}\authorrefmark{1}}

\address[1]{Beijing Sport University, Beijing 100084, China}
\address[2]{School of  Information and Communication Engineering, Beijing University of Posts and Telecommunications, Beijing 100876, China}
\address[3]{Nanjing University of Science and Technology, Nanjing 210094, China}
\address[4]{Mendeleev University of Chemical Technology of Russia, Moscow 125047, Russia}

\markboth
{Author \headeretal: Preparation of Papers for IEEE TRANSACTIONS and JOURNALS}
{Author \headeretal: Preparation of Papers for IEEE TRANSACTIONS and JOURNALS}

\corresp{Corresponding author: YUHAO ZHENG, CHANG LIU (e-mail: yuhao\_zheng@bupt.edu.cn; c.liu@bsu.edu.cn).}

\begin{abstract}
In this paper, we propose a two-stage optimization framework for secure task scheduling in satellite–terrestrial edge computing networks (STECNs). The framework jointly considers secure user association and task offloading to balance transmission delay, energy consumption, and physical-layer security. To address the inherent complexity, we decouple the problem into two stages. In the first stage, a secrecy-aware user association strategy is designed by discretizing artificial noise (AN) power ratios and identifying feasible links that satisfy secrecy constraints, resulting in a set of candidate secure associations. In the second stage, we formulate a delay-energy-aware task scheduling problem as an integer linear program and solve it using a heuristic Mayfly Algorithm (MA) to obtain low-complexity, high-quality solutions. Extensive simulation results demonstrate the effectiveness and superiority of the proposed framework in achieving secure and efficient task scheduling under dynamic satellite environments.
\end{abstract}


\titlepgskip=-21pt

\maketitle

\section{Introduction}
\label{sec:introduction}
\subsection{Background}
\PARstart{W}{ith} the rapid development of sixth-generation (6G) networks and satellite communications, the demand for seamless global connectivity and ubiquitous computing services has surged dramatically \cite{6G-background}. As terrestrial infrastructures alone cannot fulfill such demand, especially in remote, oceanic, or emergency areas, the concept of satellite-terrestrial network (STN) has been proposed. STN tightly integrate low Earth orbit (LEO) satellites and ground infrastructures to achieve wide-area coverage and service availability \cite{LEO-background}. Meanwhile, mobile edge computing (MEC) has emerged as a key enabler of low-latency and computation-intensive services by bringing computational capabilities closer to end users \cite{MEC-background}. By combining the global coverage of STN with the distributed computing ability of MEC, satellite-terrestrial edge computing network (STECN) have become a promising architecture to support intelligent task scheduling with strong spatiotemporal flexibility \cite{STECN-background}, which unlocks new opportunities for satellite-terrestrial environments.

In recent years, STECN have received increasing attention, aiming to improve computational responsiveness, real-time task processing, and overall network efficiency. Existing studies have explored various technological aspects of STECN, including network slicing \cite{STECN-1}, resource allocation \cite{STECN-2}, task offloading \cite{STECN-3}, and edge server selection \cite{STECN-4}. These works have made significant progress in addressing key performance metrics such as end-to-end latency, resource utilization, and energy consumption. However, STECN faces a multitude of unique constraints that significantly complicate the task scheduling process. First, LEO satellites are characterized by rapid orbital movement, which leads to highly dynamic and time-varying network topologies. This results in frequent handovers, rapidly changing link quality, and intermittent satellite visibility for ground users \cite{dynamic-background}. Second, due to limitations in payload capacity and power supply, each satellite is equipped with only a modest amount of computational and energy resources \cite{limited-background}. These limitations severely restrict the number of tasks a satellite can process and limit the computational intensity of the services it can support. Third, task scheduling in STECN must be optimized not only across space but also across time, while taking into account varying channel conditions and the presence of eavesdroppers. This spatiotemporal complexity is further compounded by dynamic security threats, where myopic scheduling focused only on delay or energy may increase exposure to risks or cause future resource bottlenecks.

Beyond the challenge of secure transmission, most existing research assumes a trusted or closed network environment, paying little attention to the critical issue of secure transmission, which is especially crucial in the open and dynamic space-ground scenario \cite{secure-background-1}. The broadcast nature of satellite links, combined with increasing threats from passive or active adversaries, makes task data highly vulnerable to eavesdropping and interception. This challenge is particularly significant in mission-critical and privacy-sensitive application scenarios, such as military surveillance, emergency response, remote medical diagnosis, and confidential industrial operations \cite{secure-background-2}, where user tasks may contain sensitive information and privacy-critical models. In such contexts, ignoring security considerations in task scheduling may lead to severe privacy leakage or mission failure. Therefore, STECN must support secure task offloading and association, which not only satisfies traditional performance goals but also meets physical-layer secrecy constraints. Secure task scheduling under secrecy capacity limitations remains largely unexplored in the STECN context.

\subsection{Related Works}
Recent years have witnessed growing interest in STECN, which provides low-latency and computation-intensive services in a broader coverage area \cite{STECN-background}. A variety of technological dimensions have been explored in this context. For example, Gao \textit{et al.} \cite{STECN-4} employed a hybrid approach combining K-center clustering and NSGA-II to reduce latency and energy consumption in STECN environments. Esmat \textit{et al.} \cite{STECN-1} investigated resilient network slicing techniques tailored for satellite edge computing (SEC), emphasizing service continuity through dynamic management and orchestration. Furthermore, Wei \textit{et al.} \cite{STECN-3} addressed joint slicing and offloading through a bi-level game model and distributed solution design, while Esmat \textit{et al.} \cite{STECN-5} proposed a cross-domain slicing mechanism based on restless multi-armed bandit (RMAB) processes for satellite-ground resource coordination. While these studies have provided valuable insights into resource allocation, task offloading, and service orchestration in STECN, they primarily focus on performance optimization under trusted environments, overlooking critical security issues in open and dynamic satellite-ground systems. In particular, secure task transmission and scheduling have not been systematically addressed in STECN.

Nowadays, physical-layer security (PLS) techniques have gained increasing attention for enhancing the security of satellite networks. Wyner \textit{et al.} \cite{wyner} first introduced the wiretap channel model, in which the eavesdropper receives a degraded version of the signal intended for the legitimate user. The core idea of PLS is to exploit the randomness of wireless channels to ensure secure transmission \cite{PLS-core}. Therefore, artificial noise (AN) has been proposed to transmit alongside useful information to interfere with potential eavesdroppers, while legitimate receivers can filter out this interference during demodulation \cite{AN-first,AN-intro-1,AN-intro-2}. Several studies have explored the use of AN in low Earth orbit (LEO) satellite networks. Works \cite{AN-downlink-1,AN-downlink-2} discuss the downlink transmission security using AN, which are mostly used in content delivery or signal transmission scenarios instead of considering task scheduling. Meanwhile, these works mainly focus on single satellite rather than satellite constellation. Also, there are works \cite{AN-uplink-1,AN-uplink-2} discussing the security in uplink transmission, regarding nearby satellites as eavesdroppers.  However, none of these work integrate AN techniques with task scheduling in satellite scenarios. In our paper, we consider the ground eavesdroppers instead of satellite eavesdroppers, because the unnessity and high cost for eavesdropping from other satellites. The ground eavesdropper are more realistic in our scenario.downlink secure and task scheduling transmission.

In terrestrial mobile edge computing (MEC), many existing works have already used AN to make sure secure transmission \cite{ground-work-1,ground-work-2,ground-work-3}, but only a limited number of works consider secure task scheduling and computation offloading under adversarial threats. Wu \textit{et al.} \cite{ground-secure-1} proposed a cooperative jamming-assisted computation offloading scheme that jointly optimizes service caching, transmit power, and offloading decisions to minimize delay in the presence of eavesdropping threats. Zhou \textit{et al.} \cite{ground-secure-2} designed a secure computation offloading framework for cache-assisted ultra-dense MEC networks, which minimizes energy consumption under strict delay, resource, and security constraints. Similarly, Zahed \textit{et al.} \cite{ground-secure-3} developed a green and secure MEC architecture for IoT applications, integrating edge caching, cooperative task offloading, and security service assignment to reduce both energy and potential security breach costs. While these works address security-aware in terrestrial MEC networks, but they focus more on content delivery, and seldom consider the task scheduling problem. The secure task scheduling topic is a notable absence of such research in the satellite context, where mobility, coverage variation, and resource dynamics introduce additional challenges.

\subsection{Contributions}
We propose a two-stage delay-energy-security aware scheduling framework, explicitly designed to cope with the spatiotemporal complexity of STECN. By discretizing the dynamic LEO satellite network into time slots and jointly capturing spatial variations and temporal variations, the framework enables adaptive and secure task scheduling under rapidly changing conditions. The proposed framework integrates a secure user association mechanism and an efficient task scheduling strategy. At the first stage, we formulate a secure association strategy under secrecy capacity constraints to protect users from eavesdropping threats. At the second stage, we develop a delay-energy-aware task scheduler based on mayfly algorithm (MA), enabling task scheduling under time-varying environment. The main contributions are summarized as follows:
\begin{itemize}
    \item We design the system architecture of STECN and detail the secure scheduling process. A two stage framework is developed to accurately capture dynamic states of LEO satellites and user requests across different time slots.
\item At the first stage, we formulate the user association problem under secrecy capacity constraints, and develop a secure association strategy that dynamically selects LEO satellites to ensure privacy-preserving and adaptive user access in dynamic topologies.
\item At the second stage, we focus on task scheduling under delay-energy performance. MA algorithm is designed to efficiently schedule tasks among satellites, minimizing delay and energy consumption while adapting to resource fluctuations and link variability.
\item Extensive simulations validate the proposed framework under practical LEO dynamics, demonstrating its effectiveness in reducing task delay and energy consumption while ensuring secure service delivery, compared with baseline approaches.
\end{itemize}

The structure of this paper is organized as follows. \cref{sec:system} presents the system model and problem formulation. \cref{sec:algorithm} details the proposed two-stage heuristic algorithm. \cref{sec:simulations} reports simulation results and performance evaluation. \cref{sec:conclusion} concludes the paper and outlines future research directions.

\section{System Description}
\label{sec:system}
In this section, we first describe our system model in our research.

\subsection{Network Model}
We consider a LEO satellite network that supports secure task scheduling from LEO satellites to legitimate users. Let \( \mathcal{N} = \{1, \dots, N\} \) denote the set of LEO satellites, and \( \mathcal{U} = \{1, \dots, U\} \) denote the set of legitimate users. Each LEO satellite $n\in\mathcal{N}$ can perform computational service for users. The eavesdroppers can be denoted as \( \mathcal{E} = \{1, \dots, E\} \). We assume that the LEO satellite network has secure inter-satellite links (ISLs), which enable data exchange between LEO satellites without risk of eavesdropping. Due to the highly dynamic and periodic nature of LEO satellites, the system operates in discrete time slots indexed by $t \in \mathcal{T}=\{1, 2, \ldots, T\}$. By analyzing the status of users and satellites in different time slots, the problem of time and space complexity can be effectively handled. Due to constrained computing resources onboard, each LEO satellite can execute a limited number of tasks. Let \( \mathcal{R}(t) = \{r_1(t), \dots, r_U(t)\} \) be the set of all user task requests. Task scheduling decision can be denoted by binary variable \( o_{n,u}(t) \in \{0, 1\} \), where \( o_{n,u}(t) = 1 \) indicates that LEO satellite \( n \) execute task \( r_u \) in time $t$, and 0 otherwise. The computational requirements of task \( r_u(t)\) is denoted as $c_u$.

Due to the limited beam coverage of LEO satellites, user association is constrained by the beamwidth. Let \( \theta_{\text{beam}} \) be the half beamwidth angle of the LEO satellite, and \( \theta_{n,u}(t) \) denote the central angle between the LEO satellite \( n \) and the user \( u \) in time $t$. User \( u \) can be associated with LEO satellite \( n \) only if it lies within the beam coverage region of the LEO satellite. The upper bound of the central angle can be mathematically expressed as \cite{AN-uplink-2}:
\begin{equation}
    \theta_{\text{max}} = 
    \begin{cases}
        \sin^{-1}\left( \dfrac{R_n}{R_{0}} \sin\theta_{\text{beam}} \right) - \theta_{\text{beam}}, 
        & \theta_{\text{beam}} < \sin^{-1}\left( \dfrac{R_{0}}{R_n} \right) \\
        \cos^{-1}\left( \dfrac{R_{0}}{R_n} \right),
        & \theta_{\text{beam}} \geq \sin^{-1}\left( \dfrac{R_{0}}{R_n} \right)
    \end{cases}
\end{equation}
where $R_0,R_n$ are the radius of the earth and LEO satellites. Let $\Psi_u(t)$ denotes the available set of LEO satellites whose $\theta_{n,u}(t)$ is in the range $[0, \theta_{max}]$ for user $u$. The user chooses the LEO satellite in $\Psi_u(t)$ to be the association LEO satellite $n_0$. We define $a_{n,u}(t)\in\{0,1\}$ to represent the association decision of user $u$. When user $u$ is directly associated with LEO satellite $n$ in time $t$, $a_{n,u}(t) = 1$; otherwise, $a_{n,u}(t) = 0$.

\subsection{Secure Communication Model}
We incorporate artificial noise and both large-scale and small-scale fading to characterize the communication model.

\subsubsection{Artificial Noise}
In this article, we introduce AN into the transmitted signal to enhance its secrecy. Our goal is to ensure that legitimate users can correctly receive the transmitted data, while eavesdroppers cannot successfully decode the information. Let \( P_t \) be the total transmit power allocated by the satellite, which is divided between the useful signal and the AN. Let \( \psi \) denote the information-bearing ratio, which falls within the interval [0, 1]. Then the power for the signal is \( \psi P_t \), and the power for AN is \( (1 - \psi)P_t \).

\subsubsection{Downlink Model}
We model satellite-to-ground wireless channel. Large-scale fading is primarily modeled by the free-space path loss. The large-scale path loss between LEO satellite \( n  \) and user \( u  \) is given by:
\begin{equation}
L_{n,u}(t) = \left( \frac{c}{4\pi f_c d_{n,u}(t)} \right)^2
\end{equation}
where \( c \) is the speed of light, \( f_c \) is the carrier frequency, and \( d_{n,u}(t) \) denotes the distance between LEO satellite \( n \) and user \( u \). We adopt the Shadowed-Rician (SR) fading model, which is widely used for satellite-ground communication \cite{SR}. The small scale fading coefficient is denoted as \( |h_{n,u}(t)|^2 \), where \( h_{n,u}(t) \) is a complex random variable governed by the SR distribution. Combining the effects of large-scale path loss and small-scale fading, the received signal power at user \( u \) from LEO satellite \( n \) is given by:
\begin{equation}
P_{n,u}^{\text{rec}}(t) = P_t \cdot G \cdot L_{n,u}(t) \cdot |h_{n,u}(t)|^2
\end{equation}
where \( G \) represents the combined antenna gain of the LEO satellite transmitter and ground receiver. Then, the signal-to-interference and noise ratio (SINR) at the legitimate user can be given by:
\begin{equation}
\text{SINR}_{n_0,u}(t) = \frac{\psi P_{n_0,u}^{\text{rec}}(t)}{\sum_{n\in \{\Psi_u/n_0\}} P_{n,u}^{\text{rec}}(t) + \sigma^2},
\end{equation}
where \( \sigma^2 \) is the noise power. The corresponding downlink transmission rates for the legitimate user can be expressed as:
\begin{equation}
R_{n_0,u}(t) = W_0 \log_2(1 + \text{SINR}_u(t))
\label{RTP}
\end{equation}
where $W_0$ is the available LEO satellite bandwidth. Similarly, the SINR at the  eavesdropper can be given by:
\begin{equation}
\text{SINR}_{n_0,e}(t) = \frac{\psi P_{n_0,e}^{\text{rec}}(t)}{(1 - \psi)P_{n_0,e}^{\text{rec}}(t) + \sum_{n\in \{\Psi_e/n_0\}} P_{n,e}^{\text{rec}}(t) + \sigma^2}
\end{equation}
The corresponding downlink transmission rates for the eavesdropper can be expressed as:
\begin{equation}
    R_{n_0,e}(t) = W_0\log_2(1 + \text{SINR}_e(t))
\label{ETP}
\end{equation}
Increasing the artificial noise power can effectively suppress \( \text{SINR}_{n_0,e}(t) \) and \( R_{n_0,e}(t) \), but it may also degrade \( \text{SINR}_{n_0,u}(t) \) and reduce \( R_{n_0,u}(t) \). Conversely, allocating more power to the useful signal improves \( R_{n_0,u}(t) \) but also increases \( R_{n_0,e}(t) \). Therefore, there exists a tradeoff in power allocation that needs to be carefully optimized. On this basis, the secrecy transmission rate, which is defined as the difference of the achievable rate between the legitimate transmission rate and the eavesdropping transmission rate is given by:
\begin{equation}
    R^{\text{sec}}_{n_o,u}(t) = \left[ R_{n_0,u}(t)-\max_{e\in\mathcal{E}}R_{n_0,e}(t)\right]^+
\label{RE}
\end{equation}
where $\left[x\right]^+$ is defined as $\text{max}\left\{0, x\right\}$.

\subsubsection{Inter-Satellite Link Model}
LEO satellites use lasers to achieve ISL interconnection. The relative positions of LEO satellites within the same orbital plane remain stable, and stable inter-orbital distances enable connections with nearby LEO satellites in adjacent planes \cite{adjacent}. With four inter-satellite links (ISLs), LEO satellites form a grid topology orbital network. The data transmission rate from adjacent LEO satellite $n$ to $m$ in time slot $t$ can be represented as \cite{data-transmission}:
\begin{equation}
R_{n,m}^S(t) = \frac{P_n G_n^{TX} G_m^{RX} L_{n,m}(t)}{b T_s \cdot \left( \frac{E_b}{N_0} \right)\cdot M},
\end{equation}
where $L_{n,m}(t)$ is the free space loss between LEO satellite $n$ and $m$ in time slot $t$. \( b \) is the Boltzmann constant, \( T_s \) is the total system noise temperature, \( \left( \frac{E_b}{N_0} \right)\) is the required received energy per bit relative to noise density, \( M \) is the link margin. Due to we do not regard LEO satellites as eavesdroppers, we assume the ISLs are secure during data transmission. To mitigate interference from multiple requests on the wireless channels, we use Orthogonal Frequency Division Multiplexing (OFDM) to assign a unique subchannel for each request and equally divide the total bandwidth among all subchannels. This ensures orthogonality and prevents interference between the subcarriers.

\subsection{Delay Model}
The average delay comprises the data transmission delay, the waiting delay at the data buffer, and the execution delay. The data transmission delay is determined by data size and achievable transmission rate. Let \( \zeta_u(t) \) denote the size of the result data of task $r_u(t)$, then the data transmission delay is:
\begin{equation}
D_{u,n}^{\text{trans},m}(t) = \frac{\zeta_u(t)}{R_{n_0,u}(t)} + \sum_{p,q\in e_{n_0,m}(t)}{\frac{o_{m,u}(t)\zeta_u(t)}{R_{p,q}(t)}}
\end{equation}
where $e_{n_0,m}(t)$ is the transmission path in the ISL. we employ the Floyd algorithm to determine the transmission path $e_{n_0,m}(t)$. If the task is scheduled to LEO satellite $m$, the data transmission delay comprises the downlink transmission delay of association LEO satellite $n_0$, and the ISL transmission delay from LEO satellite $m$ to $n_0$ (if exists).

The tasks execution delay is predominantly linked to the computing resources needed for the task and the computing capacities of the LEO satellites. The execution delay can be computed as follows:
\begin{equation}
D_{u,n}^{\text{comp},m} = \frac{o_{m,u}(t)\zeta_u(t)c_u(t)}{f_{m}}
\end{equation}
where \( f_m \) denotes the computing resources (CPU cycles per second) of the LEO satellite $m$.

When request \( r_u(t) \) arrives at the data queue of satellite \( m \), the computing task needs to wait for processing. We assume that \( Q_{m,u}(t) \) represents the collection of requests prior to request \( r_u(t) \) in the queue of LEO satellite \( m \) at time \( t \). Once all the previous content requests of \( r_u(t) \) have been processed, \( r_u(t) \) will be processed at LEO satellite \( m \). The waiting time of the task is determined by the execution time of all existing computing tasks in the queue. The queuing delay caused by the previously accumulated tasks for request \( r_u(t) \) can be expressed as:
\begin{equation}
D_{u,n}^{\text{queue},m}(t) = \sum_{r_d(t) \in Q_{v,u}(t)} \frac{o_{m,d}(t) \zeta_d(t) c_d(t)}{f_m},
\end{equation}
where \( r_d(t) \) represents a task request in the previous request queue \( Q_{v,u}(t) \), \( o_{m,d}(t) \) denotes the scheduling decision for task request \( r_d(t) \), \( \zeta_d(t) \) indicates the data size of task request \( r_d(t) \), and \( c_d(t) \) represents the computational requirements of task request \( r_d(t) \). Then, the average delay is thus given by:
\begin{equation}
  \begin{split}
    D_{u,n}(t) = \sum_{n \in \mathcal{N}}  \Big( D_{u,n}^{\text{trans},m}(t) + D_{u,n}^{\text{comp},m}(t) +D_{u,n}^{\text{queue},m}(t)\Big).
  \end{split}
\end{equation}
\begin{figure*}[t]
\centering
\includegraphics[width=0.8\textwidth]{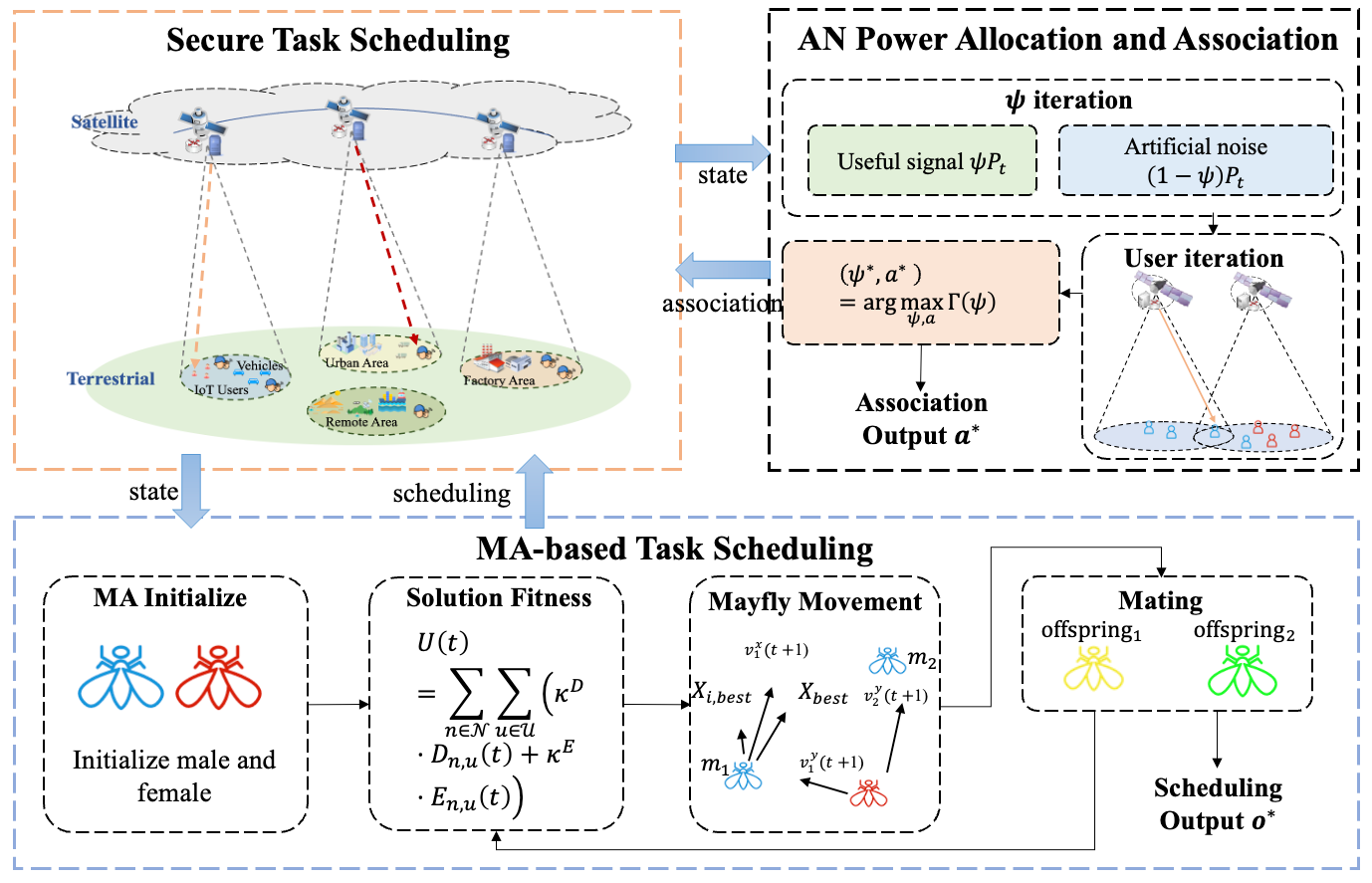}
\caption{Proposed two-stage heuristic Algorithm.}
\label{fig:system}
\end{figure*}
\subsection{Energy Consumption Model}
The energy consumption comprises the energy consumption for transmission and execution. The transmission energy consumption used to transmit task $r_u$ to LEO satellite $m$ can be expressed as follows:
\begin{equation}
  E_{n,u}^{\text{trans},m}(t) = P_t D_{n,u}^{\text{trans},m}(t) .
\end{equation}

The execution energy consumption of LEO satellite \( m \) for executing task $r_u$ can be given by:
\begin{equation}
  E_{n,u}^{\text{comp},m}(t) = \varepsilon (f_{m})^3 D_{n,u}^{\text{comp},m}(t),
\end{equation}
where \( \varepsilon (f_{m})^3 \) is the function of power and computing capability \cite{power}. Then, the average energy consumption is computed as:
\begin{equation}
  \begin{split}
    E_{n,u}(t) = \sum_{m \in \mathcal{N}} \Big( E_{n,u}^{\text{trans},m}(t) + E_{n,u}^{\text{comp},m}(t) \Big).
  \end{split}
\end{equation}

\subsection{Problem Formulation}
Due to the decision variables are defined per time slot to explicitly account for spatial and temporal dynamics, the optimization can be carried out properly under the inherent spatiotemporal complexity of STECN. Our objective is to design an efficient scheduling and user association strategy that jointly minimizes task delay and energy consumption while ensuring a minimum level of communication secrecy. Specifically, we formulate a constrained optimization problem where the goal is to minimize the total system cost composed of delay and energy terms, subject to secrecy rate and resource capacity constraints. The problem is expressed as:
\begin{equation}
\min_{a_{n,u},\ o_{n,u},\psi} \quad \sum_{t\in \mathcal{T}}\sum_{n\in\mathcal{N}}\sum_{u\in\mathcal{U}}\big(\kappa^D\cdot D_{n,u}(t) + \kappa^E\cdot E_{n,u}(t)\big)\tag{P1}\label{eq:P1}
\end{equation}
\begin{equation}
\text{subject to} \quad
\begin{cases}
\kappa^D+\kappa^E=1,  \\
R^{\text{sec}}_{n_0,u}(t) \geq \epsilon, & \forall u\in\mathcal{U}, \\
\psi \in (0, 1), & \forall n\in\mathcal{N}.
\end{cases}
\tag{C1}
\label{constrain}
\end{equation}
where \( \kappa^D, \kappa^E \) are performance gain coefficients. In the constraints, we need to ensure that the secrecy offloading rate must exceed the secrecy requirement to ensure secure computation offloading. Meanwhile, the designed AN must fall within the interval [0, 1].

\section{Proposed Two-Stage Heuristic Algorithm}
\label{sec:algorithm}
This section introduces a novel framework to address the problem of secure user association and task scheduling in two stages.

\subsection{Problem Analysis and Decomposition}
The original optimization problem \ref{eq:P1} jointly considers user association, task scheduling, and AN power configuration to minimize the delay and energy consumption while ensuring secure communication. However, the problem is formulated as a mixed-integer nonlinear program with strong coupling among binary user association decision $a_{n,u}$, task scheduling decision $o_{n,u}$, and continuous AN power ratio $\psi$. The combination of discrete and continuous variables, dynamic task arrivals, and fluctuating satellite resources makes real-time global optimization intractable for STECN. To address this challenge, we adopt a two-stage decomposition approach that decouples the secure communication structure from the scheduling decision, thereby reducing computational complexity and enabling scalable. The overall procedures are shown in \cref{fig:system} Specifically, motivated by the spatiotemporal complexity of STECN which makes real-time joint optimization of security, delay, and energy challenging, we decompose the original problem into two sequential subproblems:
\begin{itemize}
    \item First Stage: Joint AN power allocation and user association. In this stage, we aim to construct a secure communication topology by determining the user-satellite association and AN power ratio that maximize the achievable secrecy rate, under link dynamics and satellite mobility
    \item Second Stage: Delay-energy-aware task scheduling. Based on the secure associations obtained in first stage, we perform task scheduling under delay and energy constraints, using heuristic search techniques to allocate resources efficiently under satellite computation and bandwidth limitations.
\end{itemize}

This decomposition allows each subproblem to be handled independently using tailored heuristics, while preserving the overall system objective of secure and efficient task offloading in dynamic STECN environments. The following subsections detail the design and implementation of each stage.

\subsection{First Stage: Joint AN Power Allocation and User Association}
Achieving strict physical-layer secrecy (\( R^{\text{sec}}_{n_0,u} \geq \epsilon \)) for all users in STECN is often infeasible due to the randomness of eavesdropper distribution and dynamic link conditions. Simulation results show that enforcing such a hard constraint leads to frequent failure in user association. Therefore, we relax the constraint and instead maximize the number of users meeting the secrecy threshold under varying AN power ratios.

We discretize the AN power allocation ratio \( \psi \) into a finite candidate set \( \{0, \Delta, 2\Delta, \dots, 1\} \), where \( \Delta \) is a tunable step size controlling the granularity of the search space. For each candidate \( \psi \), we evaluate the achievable secrecy rate \( R^{\text{sec}}_{n,u}(\psi) \) for each user-satellite pair \( (u,n) \), and attempt to find an association such that user \( u \) is connected to a satellite \( n \in \Phi_u \) satisfying the secrecy condition:
\begin{equation}
    R^{\text{sec}}_{n,u}(\psi) = R_{n,u}(\psi) - \max_{e \in \mathcal{E}} R_{n,e}(\psi) \geq \epsilon,
\end{equation}
where \( R_{n,u}(\psi) \) and \( R_{n,e}(\psi) \) denote the achievable rates of the legitimate user and the strongest eavesdropper, respectively. We define the reliable transmission rate (RTP) under a given AN power ratio \( \psi \) as the proportion of users that can be securely served:
\begin{equation}
    \Gamma(\psi) = \frac{1}{U} \sum_{u \in \mathcal{U}} \mathbb{D} \left\{ \exists\, n \in \Phi_u : R^{\text{sec}}_{u,n}(\psi) \geq \epsilon \right\},
    \label{eq:reliable_transmission_rate}
\end{equation}
where \( \mathbb{D}\{\cdot\} \) is the indicator function. Our optimization objective at this stage becomes:
\begin{equation}
(\psi^*, a^*) = \arg\max_{\psi, a} \ \Gamma(\psi),\tag{P2}\label{eq:P2}
\end{equation}
where \( a = \{a_u\}_{u \in \mathcal{U}} \) denotes the user association decision. By jointly searching over \( \psi \) and user associations \( a \), we construct a secure communication topology that is most effective under the current conditions. 

Algorithm~\ref{alg:association} outlines the detailed process. For each candidate \( \psi \), we greedily associate users to satellites offering the highest achievable secrecy rates and record the secure association count \( \Gamma(\psi) \). If the current configuration yields a higher count than previous attempts, the algorithm updates the optimal pair \( (\psi^*, a^*) \). This stage thus produces both the optimal AN configuration and the corresponding secure association strategy, which are passed to the next stage for delay-energy-aware task scheduling.

\begin{algorithm}[t]
\caption{Joint AN Power Allocation and User Association}
\label{alg:association}
\begin{algorithmic}[1] 
\renewcommand{\algorithmicrequire}{\textbf{Input:}}
\renewcommand{\algorithmicensure}{\textbf{Output:}}

\Require LEO satellite set \(\mathcal{N}\), legitimate user set \(\mathcal{U}\), eavesdropper set \(\mathcal{E}\), secrecy rate threshold \(\epsilon\)

\State Initialize $a^\star \leftarrow \varnothing$, $\psi^\star \leftarrow 0$, $\Gamma_{\max}\leftarrow 0$
\ForAll{$\psi \in \{0,\Delta,2\Delta,\dots,1\}$}
    \State Initialize $a \leftarrow \varnothing$, $\Gamma(\psi)\leftarrow 0$
    \ForAll{$u \in \mathcal{U}$}
        \State Initialize $R^{\text{sec}}_{\max}\leftarrow -\infty$, \,$n^\star\leftarrow\varnothing$
        \ForAll{LEO satellites $n \in \Phi_u$}
            \State compute user $R_{n,u}(\psi)$ by Eq.~(\ref{RTP})
            \State compute eavesdropper $R_{n,e}(\psi)$ by Eq.~(\ref{ETP})
            \State compute secrecy $R^{\text{sec}}_{n,u}(\psi)$ by Eq.~(\ref{RE})
            \If{$R^{\text{sec}}_{n,u}(\psi) > R^{\text{sec}}_{\max}$}
                \State Update $R^{\text{sec}}_{\max}\leftarrow R^{\text{sec}}_{n,u}(\psi)$; \,$n^\star\leftarrow n$
            \EndIf
        \EndFor
        \State Assign $a_u \leftarrow n^\star$
        \If{$R^{\text{sec}}_{\max} \ge \epsilon$}
            \State Increment $\Gamma(\psi)\leftarrow \Gamma(\psi) + 1$
        \EndIf
    \EndFor

    \If{$\Gamma(\psi) > \Gamma_{\max}$}
        \State Update $\Gamma_{\max}\leftarrow \Gamma(\psi)$; \,$a^\star \leftarrow a$; \,$\psi^\star \leftarrow \psi$
    \EndIf
\EndFor
\Ensure Optimal association \(a^\star\) and AN power ratio \(\psi^\star\)
\end{algorithmic}
\end{algorithm}

\subsection{Second Stage: MA-Based Delay-Energy-Aware Task Scheduling}
After determining the secure user association and AN power ratio in the first stage, we proceed to optimize the task scheduling strategy. In this stage, each individual mayfly encodes a candidate offloading decision \( o = \{o_{n,u}\}_{n\in\Psi_u, u\in\mathcal{U}} \), where \( o_{n,u} = 1 \) indicates that user task is assigned to satellite \( n \). The goal is to minimize a weighted sum of delay and energy consumption across all users:
\begin{equation}
\min_{o} \sum_{n \in \mathcal{N}} \sum_{u \in \mathcal{U}} \left( \kappa^D \cdot D_{n,u} + \kappa^E \cdot E_{n,u} \right),
\tag{P3}
\label{eq:P3}
\end{equation}
where \( \kappa^D + \kappa^E = 1 \), controlling the balance between delay and energy concerns. Given the combinatorial nature of the binary scheduling variables and resource coupling across satellites, this problem is NP-hard and difficult to solve exactly in real time. To this end, we adopt the bio-inspired Mayfly Algorithm (MA) \cite{ma} to efficiently explore the solution space. Compared with traditional heuristics, MA incorporates both swarm-based movement (via velocity-position updates) and evolutionary mating, which improves convergence speed and avoids premature stagnation.

By modeling each scheduling strategy as a mayfly individual, we iteratively evolve the population through attraction-based movement, nuptial dance adjustment, and mating crossover. This enables the framework to adaptively balance global search and local refinement. The full MA-based scheduling procedure is described below.

$\bullet$ \textit{Initialization of Candidate Solutions:} We define the male and female populations for task scheduling as:
\begin{equation}
X = [X^1, \dots, X^I]^T, \quad Y = [Y^1, \dots, Y^I]^T,
\end{equation}
where $X$ denotes the male mayfly population, $Y$ denotes the female mayfly population. The number of male and female is the same. The initial positions and velocities are randomly initialized.

$\bullet$ \textit{Fitness Evaluation:} Each individual’s fitness is computed based on the objective function in \ref{eq:P3}, producing:
\begin{equation}
\text{Fit}^x = [\text{Fit}^x_1, \dots, \text{Fit}^x_I], \quad \text{Fit}^y = [\text{Fit}^y_1, \dots, \text{Fit}^y_I].
\end{equation}
We define the global best fitness among males as:
\begin{equation}
\text{Fit}^x_{\text{best}}(t) = \min_{i} \text{Fit}^x_i(t),
\end{equation}
where $\text{Fit}^x_i (t)$ is the fitness value of male mayfly $i$ at time $t$. Let $x_{i,\text{best}}$ denote the position of historical best fitness value $\text{Fit}^x_{i,best}(t)$ of male individual $X_i$. Let $x_{\text{best}}$ denote the global best position of male individuals.

$\bullet$ \textit{Male Mayfly Movement:} Male mayflies gather in swarms, and their position updates are influenced by neighboring individuals. The position update of each male mayfly is given by:
\begin{equation}
x_i(t+1) = x_i(t) + v_i(t+1),
\label{eq:position_update}
\end{equation}
where the velocity update follows:
\begin{align}
v_{i}^d(t+1) &= v_{i}^d(t) + a_1 e^{-\beta r_p^2}(x_{i,\text{best}}^d - x_i^d) \notag \\
&\quad + a_2 e^{-\beta r_g^2}(x_{\text{best}}^d - x_i^d).
\label{eq:velocity_update}
\end{align}
where $a_1$ and $a_2$ are positive attraction constants, $\beta$ is the visibility coefficient of the mayfly. $r_p$ and $r_g$ represent the cartesian distances between $x_i$ and both $x_{i,\text{best}}$ and $x_{\text{best}}$, respectively. Additionally, the best males perform a nuptial dance by adding random movement:
\begin{equation}
v_{i}^d(t+1) = v_{i}^d(t) + k \cdot r, \quad r \in [-1,1].
\label{eq:nuptial_dance}
\end{equation}
where $k$ is the nuptial dance distance coefficient, and $r$ is a random number in the range $[-1,1]$.

\begin{algorithm}[t]
\caption{MA-based Delay-Energy-Aware Task Scheduling Algorithm}
\label{alg:ma_task_scheduling}
\begin{algorithmic}[1]
\renewcommand{\algorithmicrequire}{\textbf{Input:}}
\renewcommand{\algorithmicensure}{\textbf{Output:}}

\Require Task set $\mathcal{T}$, available LEO satellites $\mathcal{N}$, secure association vector $a^\star$, system parameters

\State Initialize MA parameters
\State Randomly initialize male mayfly population $X$, female population $Y$, and velocities $v$

\For{$\text{iter} = 1$ to $\text{Iter}_{\max}$}
    \ForAll{male mayfly $X_i \in X$}
        \State Compute fitness $\text{Fit}_i(t)$ using \ref{eq:P3}
        \State Update global, personal best $\text{Fit}_{\text{best}}^x(t)$, $\text{Fit}_{i,\text{best}}^x(t)$
        \State Update position, velocity using \cref{eq:position_update}, \cref{eq:velocity_update}
        \If{$X_i$ is global best}
            \State Apply nuptial dance update using \cref{eq:nuptial_dance}
        \EndIf
    \EndFor

    \ForAll{female mayfly $Y_i \in Y$}
        \State Update position and velocity using \cref{eq:female_position_update}
        \State Update velocity using \cref{eq:female_velocity_update}
        \State Generate offspring using \cref{eq:offspring}
    \EndFor
\EndFor

\State \Return $o^\star$

\Ensure Optimized task scheduling decision $o^\star$

\end{algorithmic}
\end{algorithm}

$\bullet$ \textit{Female Mayfly Movement:} Female mayflies are attracted to males based on fitness and distance, updating position as:
\begin{equation}
y_i(t+1) = y_i(t) + v_i(t+1),
\label{eq:female_position_update}
\end{equation}
and adjusting velocity via:
\begin{equation}
v_{i}^d(t+1) =
\begin{cases} 
v_{i}^d(t) + a_2 e^{-\beta r_{mf}^2}(x_{i}^d - y_i^d), & \text{if } \text{Fit}_i^y > \text{Fit}_i^x, \\
v_{i}^d(t) + fl \cdot r, & \text{otherwise}.
\end{cases}
\label{eq:female_velocity_update}
\end{equation}
where $r_{mf}$ is the cartesian distance between male and female mayflies. The parameter $fl$ is the random walk coefficient.

$\bullet$ \textit{Mating and Offspring Generation:} Each male-female pair undergoes crossover to produce two offspring:
\begin{align}
\text{offspring}_1^i &= L \cdot X^i + (1 - L) \cdot Y^i, \\
\text{offspring}_2^i &= (1 - L) \cdot X^i + L \cdot Y^i,
\label{eq:offspring}
\end{align}
where $L$ is a random scalar in $[0,1]$.

$\bullet$ \textit{Termination and Output:} The algorithm terminates after a maximum number of iterations. The best individual from the final population is returned as the optimal task scheduling decision.

\subsection{Two-Stage Optimization Framework}
We propose a two-stage heuristic optimization framework to address the delay-energy-security-aware scheduling problem in STECN. In the first stage, the algorithm iteratively searches over a set of discretized AN power ratios and identifies the secure user association strategy that maximizes the number of users meeting the secrecy rate constraint. In the second stage, based on the selected secure association, we apply a mayfly algorithm (MA) to optimize task scheduling decisions with respect to delay and energy consumption. This hierarchical design decouples the security and resource objectives, allowing efficient and scalable optimization in dynamic satellite-edge environments.

The overall complexity of the proposed two-stage framework includes secure user association with AN power allocation and task scheduling via the mayfly algorithm. In the first stage, we evaluate $\frac{1}{\Delta}$ candidate AN ratios, each involving user-to-satellite association with complexity $\mathcal{O}(U )$, resulting in $\mathcal{O}(\frac{1}{\Delta} \cdot U)$. In the second stage, the MA-based scheduler operates on the selected association with complexity $\mathcal{O}(\text{Iter}_{\max} \cdot I)$, where $I$ is the population size and $\text{Iter}_{\max}$ is the iteration limit. Therefore, the total computational complexity is $\mathcal{O}(\frac{1}{\Delta} \cdot U + \text{Iter}_{\max} \cdot I)$, which remains tractable under typical parameter settings. Fig.\ref{fig:system} and Algorithm \ref{alg:twostage} show the overall procedures of the algorithm.

\begin{algorithm}[t]
\caption{Two-Stage Delay-Energy-Security-Aware Scheduling Framework}
\label{alg:twostage}
\begin{algorithmic}[1]

\renewcommand{\algorithmicrequire}{\textbf{Input:}}
\renewcommand{\algorithmicensure}{\textbf{Output:}}

\Require Time period $\mathcal{T}$; legitimate user set $\mathcal{U}$; LEO satellite set $\mathcal{N}$; AN ratio set $\{\psi\}$

\ForAll{$t \in \mathcal{T}$}

    \State Update topology, channel, user task arrivals states

    \State \textit{// Stage 1: Secure Association}
    \State $a_t^\star, \psi_t^\star \gets$ Algorithm \ref{alg:association}
    
    \State \textit{// Stage 2: Task Scheduling}
    \State $o_t^\star \gets$ Input $a_t^\star, \psi_t^\star$ into Algorithm \ref{alg:ma_task_scheduling}

    \State Execute scheduling and association $(a_t^\star, o_t^\star)$ for slot $t$

\EndFor

\State \Return $\{(a_t^\star, o_t^\star)\}_{t\in\mathcal{T}}$

\Ensure Scheduling and association decision $\{(a_t^\star, o_t^\star)\}_{t\in\mathcal{T}}$

\end{algorithmic}
\end{algorithm}

\section{Simulations}
\label{sec:simulations}
In this section, we perform extensive simulations to evaluate the performance of our proposed algorithm.

\subsection{Simulation Setup}
During the simulation process, we used the Python to construct our software environment. The experimental parameters are determined by considering the insights and methodologies outlined in the previous works: \cite{data-transmission,simulation-1}. We utilize satellite tool kit (STK) to acquire latitude and longitude data for the LEO satellites. We consider a LEO satellites constellation consisting of 20 satellite nodes with an orbital altitude of 750km. The orbit inclination is set at 58.5 degrees. Specifically, users and eavesdroppers are randomly generated within the satellite coverage area of [23.7°N–36.6°N, 94.5°E–129.1°E] during the first time slot. This coverage is formed by a constellation of 20 LEO satellites deployed over 5 adjacent orbital planes, with each plane hosting 4 satellites. The simulation window starts from 2025-07-01 04:00:00, and each time slot spans 5 seconds. We set the beamwidth of satellites to 40 degrees. We set the transmit power of satellite nodes to 5W, and the transmit and receive antenna gains are set to 12dB. The overall system noise temperature is 25dBK. The ratio of the necessary received energy per bit to the noise density is established at 9.6dB, with a designated link margin of 2500km. The available satellite bandwith is set to 1MHz. We set different number of active legitimate users and eavesdroppers in each timeslot. $\Delta$ in the first stage is set to 0.05. The task generating size is ranging from 10MB to 20MB. The computational workload for processing each bit computing task is modeled by a uniform distribution within [300, 500] CPU cycles.

The simulation is performed on the following schemes. For the various heuristic algorithms used in the second task scheduling stage, the population size is 30 and the maximum number of iterations is 200. We conducted a series of experiments to obtain accurate and reliable results. To evaluate the effectiveness of our proposed scheme, we compared it with the following baseline schemes:
\begin{itemize}
    \item {Random scheme:} In the first stage, random scheme means that users associate available LEO satellite randomly. In the second stage, random schemes means that user tasks randomly schedule to LEO satellites.
    \item {Greedy scheme:} In the first stage, user choose the LEO satellite which has the minimum $\theta_{n,u}$ to associate. In the second stage, user schedules task to the nearest LEO satellite.
    \item {Without AN scheme:} In the first stage, LEO satellites do not use AN for transmission, and user associates with available LEO satellite with better $R^\text{sec}_{n,u}$.
    \item {PSO scheme:} In the second stage, we use particle swarm optimization (PSO) \cite{pso} algorithm to obtain the scheduling decision.
    \item     {GA scheme:} In the second stage, we use genetic algorithm (GA) \cite{ga} to obtain the scheduling decision.
\end{itemize}

\subsection{Simulation Results}
\begin{figure}[!t]
\centering
\includegraphics[width=0.5\textwidth]{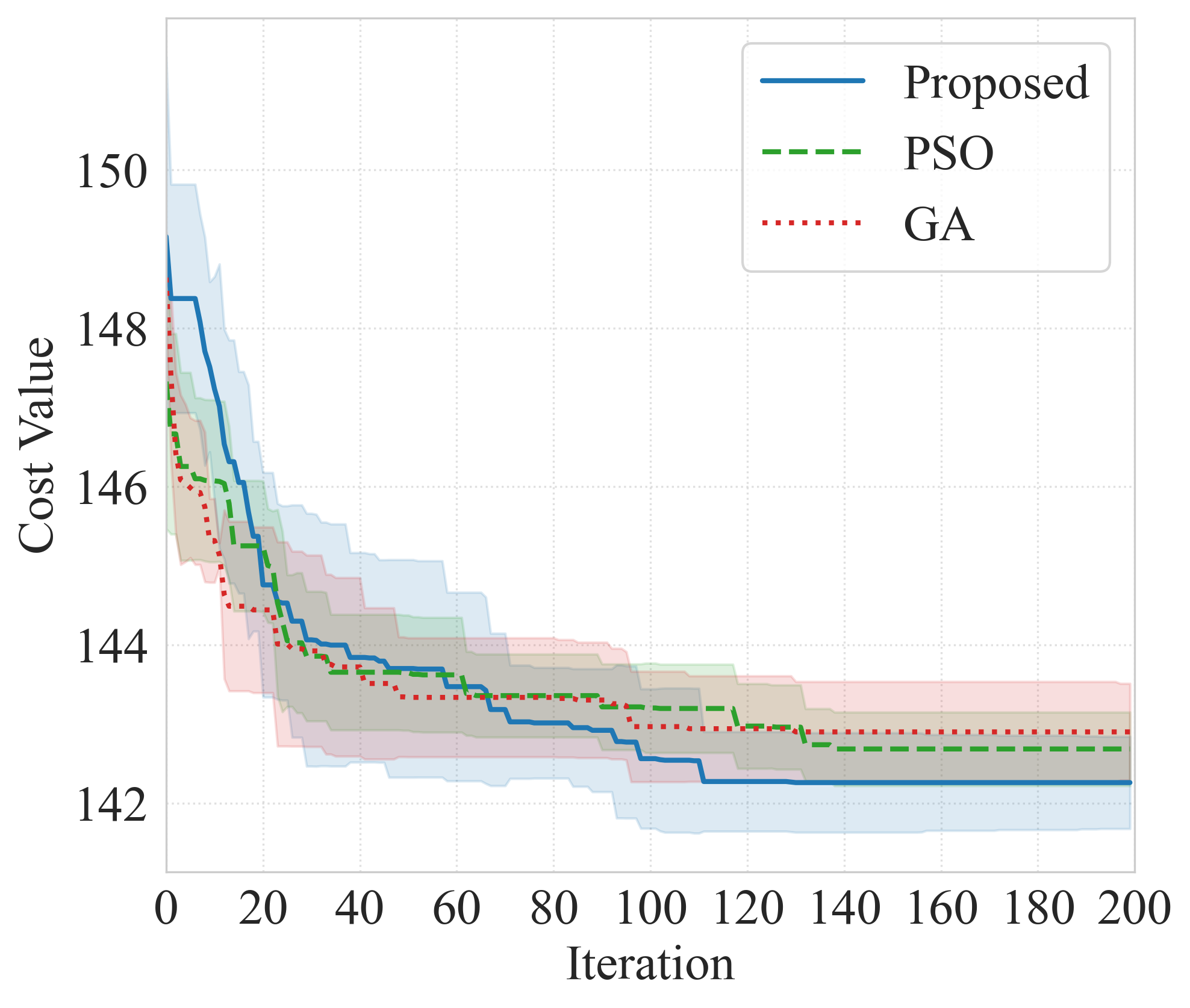}
\caption{Task scheduling converge performance of different schemes.}
\label{Converge}
\end{figure}

\begin{figure*}
	\centering
	\subfloat[\textnormal{RTP performace}] 
	{
			\begin{minipage}[b]{.31\textwidth}
					\centering
					\includegraphics[width=\textwidth]{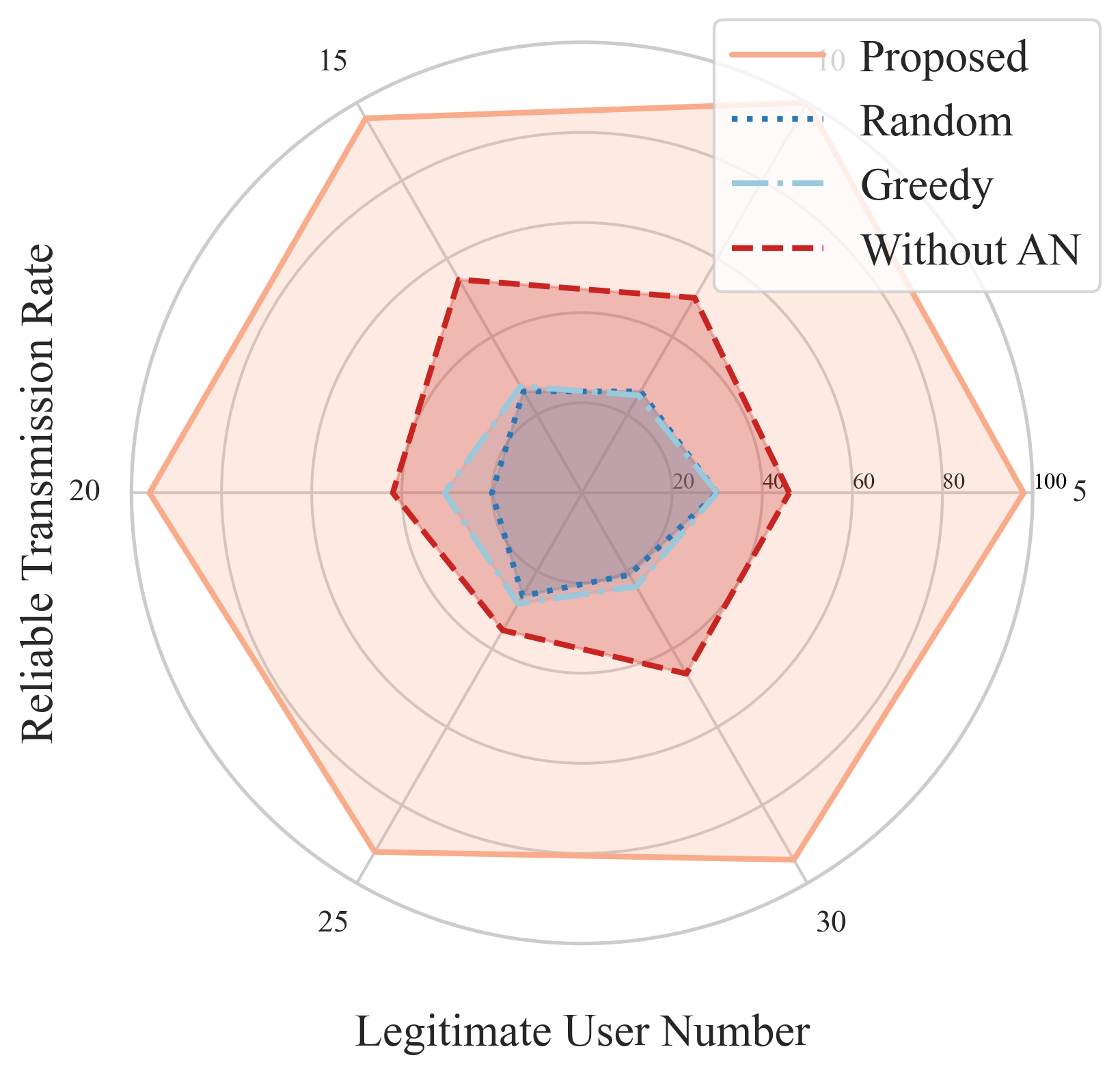}
					\label{Secure-Legitimate}
                    \vspace{-5mm} 
			\end{minipage}
		}
	\subfloat[\textnormal{Average delay performance}] 
	{
			\begin{minipage}[b]{.31\textwidth}
					\centering
					\includegraphics[width=\textwidth]{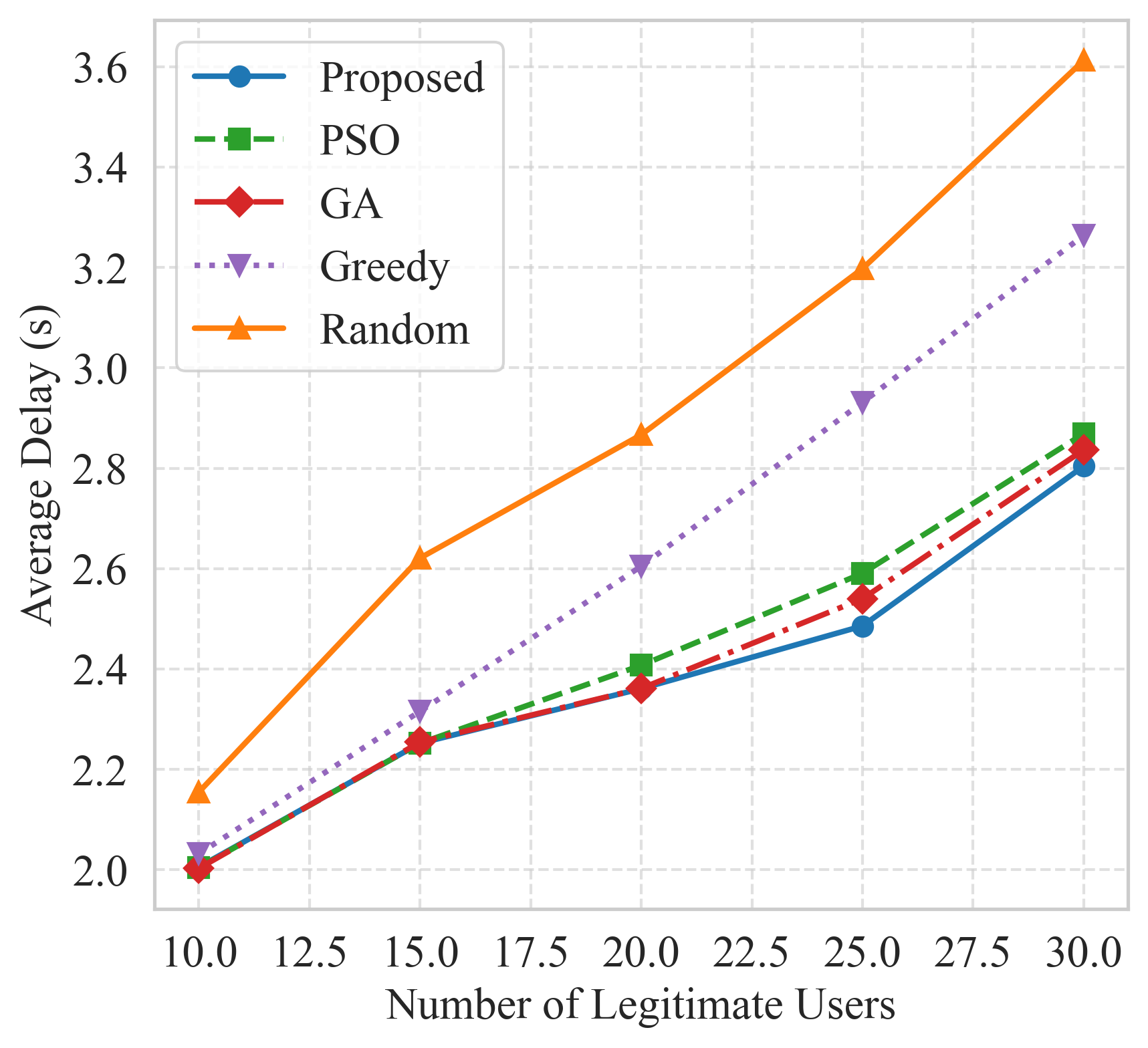}
					\label{Legitimate_Delay}
                    \vspace{-5mm} 
			\end{minipage}
		}
    \subfloat[\textnormal{Average energy consumption performance}] 
	{
			\begin{minipage}[b]{.31\textwidth}
					\centering
					\includegraphics[width=\textwidth]{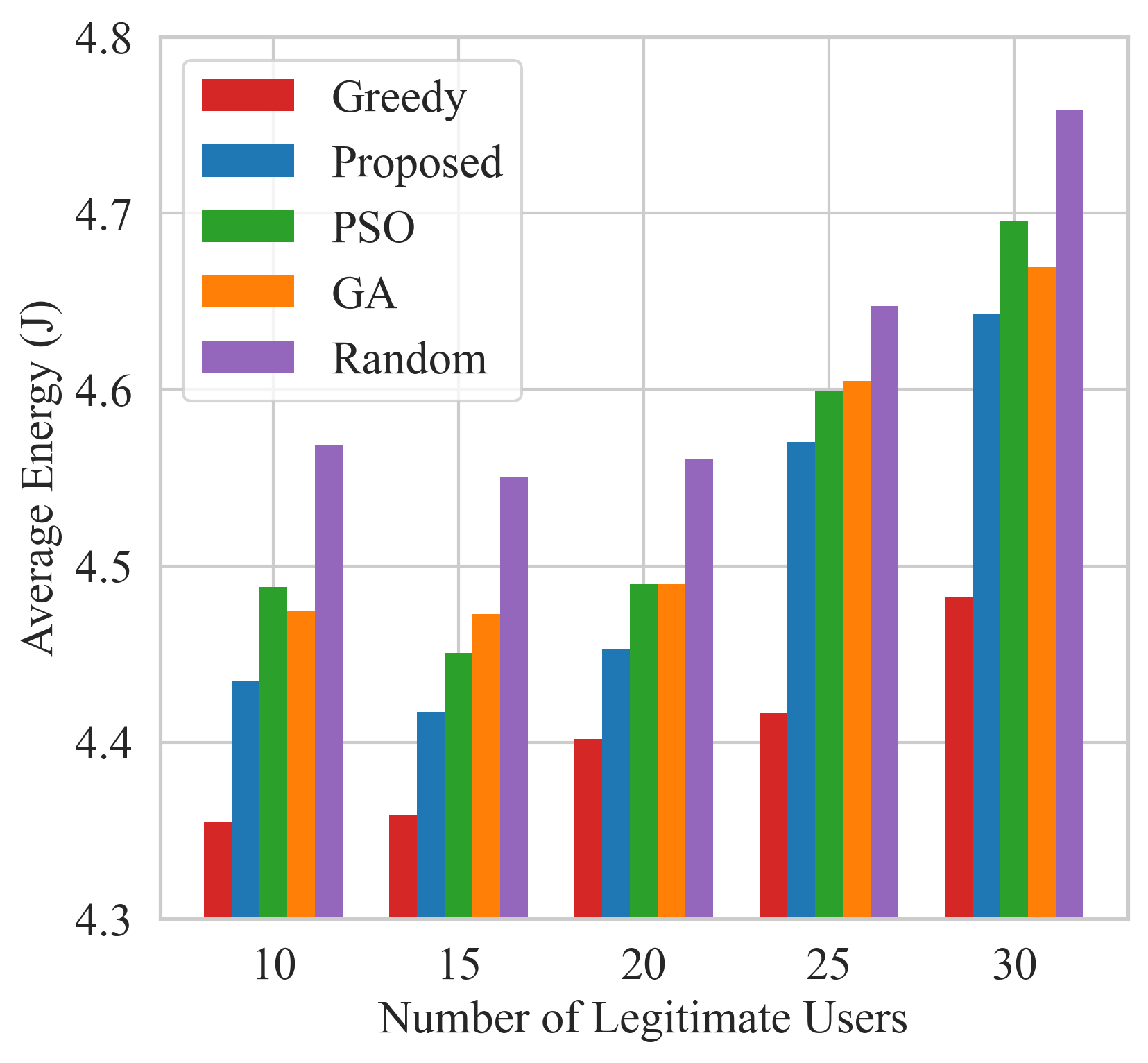}
					\label{Legitimate_Energy}
                    \vspace{-5mm} 
			\end{minipage}
	}

	\caption{Performance of different schemes under different number of Legitimate users.}
    \label{DRL_Performance}
\end{figure*}

\begin{figure*}
	\centering
	\subfloat[\textnormal{RTP performance}] 
	{
			\begin{minipage}[b]{.31\textwidth}
					\centering
					\includegraphics[width=\textwidth]{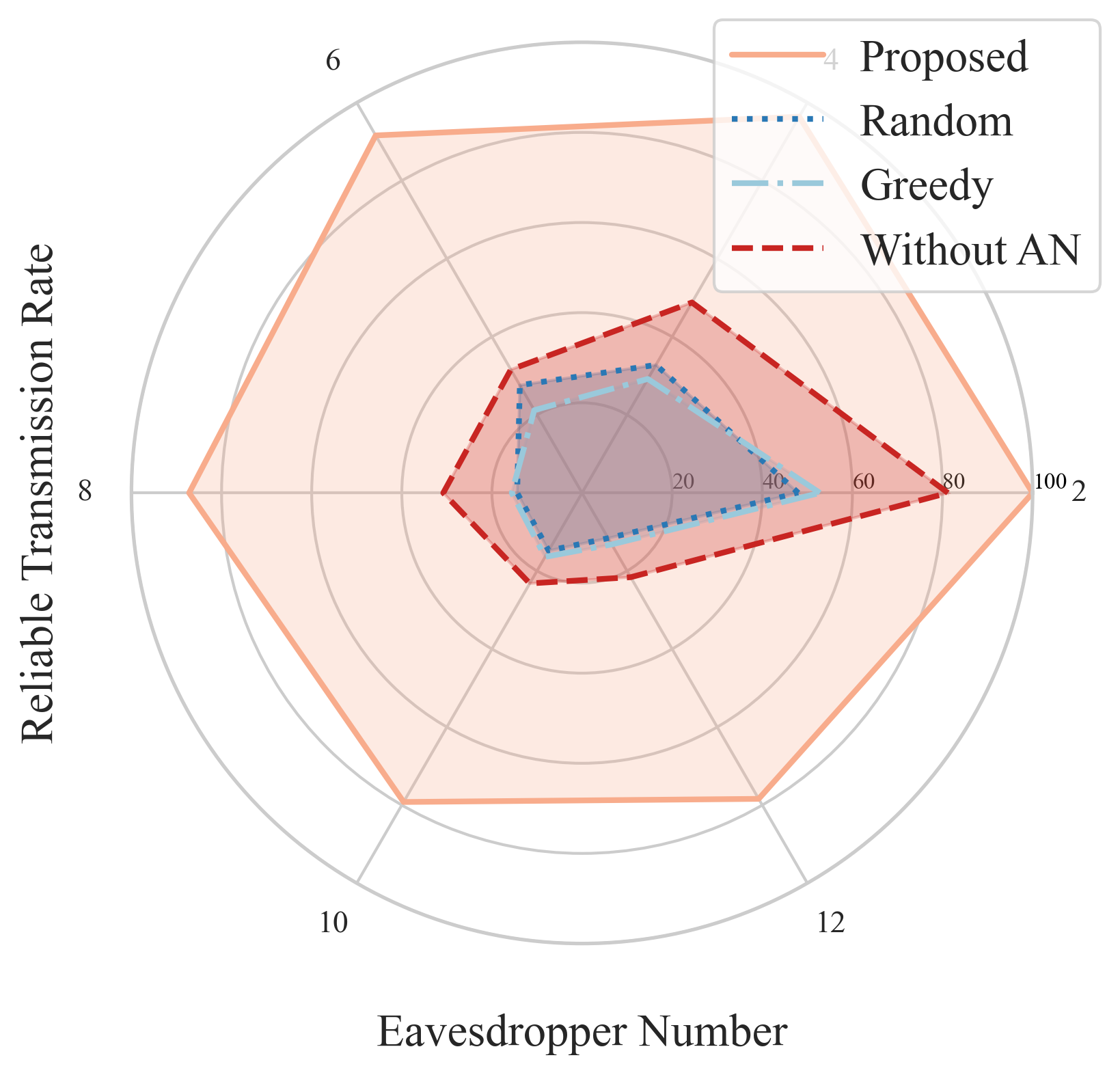}
					\label{Secure-Eavesdropper}
                    \vspace{-5mm} 
			\end{minipage}
		}
	\subfloat[\textnormal{Average delay performance}] 
	{
			\begin{minipage}[b]{.31\textwidth}
					\centering
					\includegraphics[width=\textwidth]{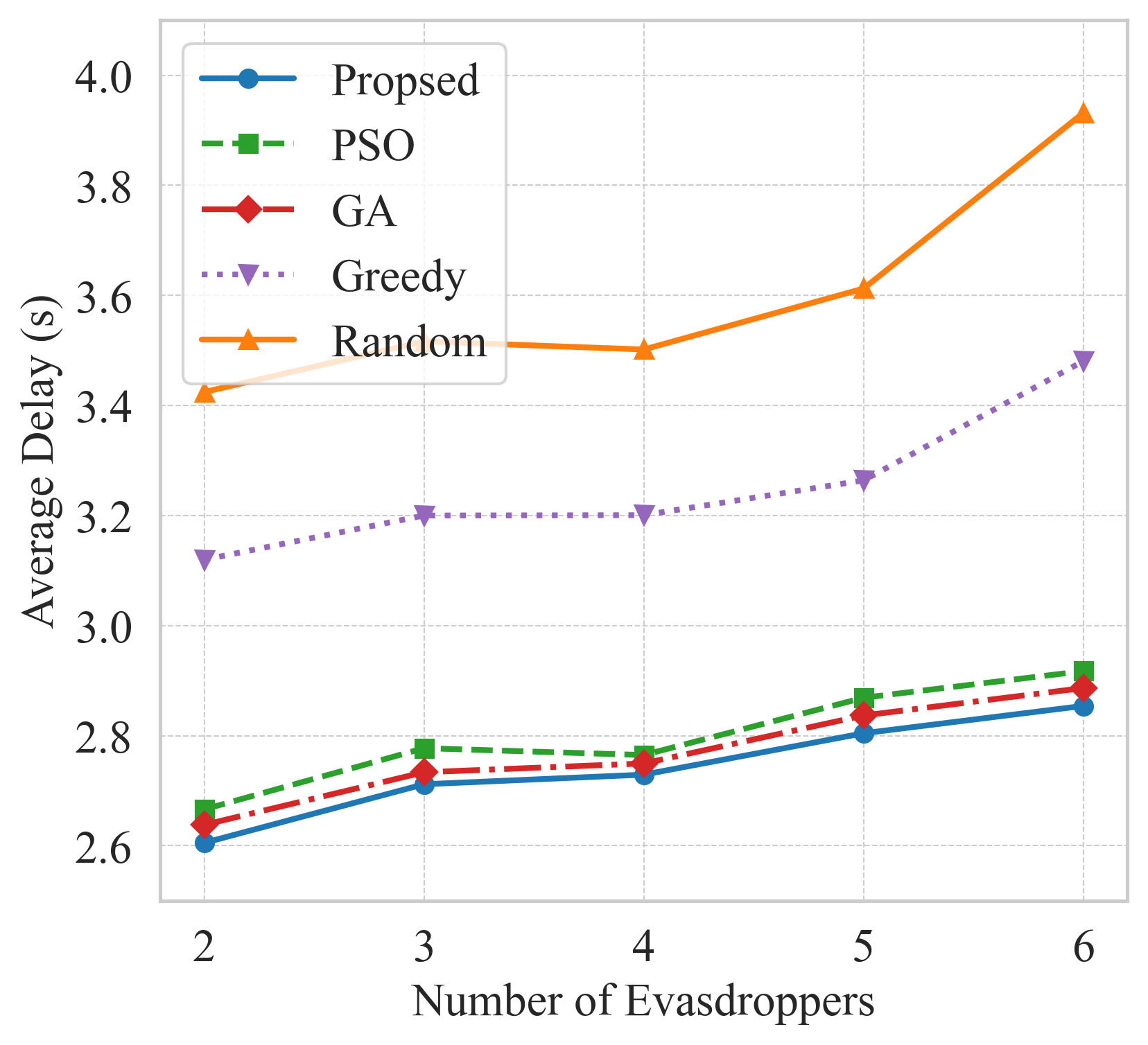}
					\label{Eavesdropper_Delay}
                    \vspace{-5mm} 
			\end{minipage}
		}
    \subfloat[\textnormal{Average energy consumption performance}] 
	{
			\begin{minipage}[b]{.31\textwidth}
					\centering
					\includegraphics[width=\textwidth]{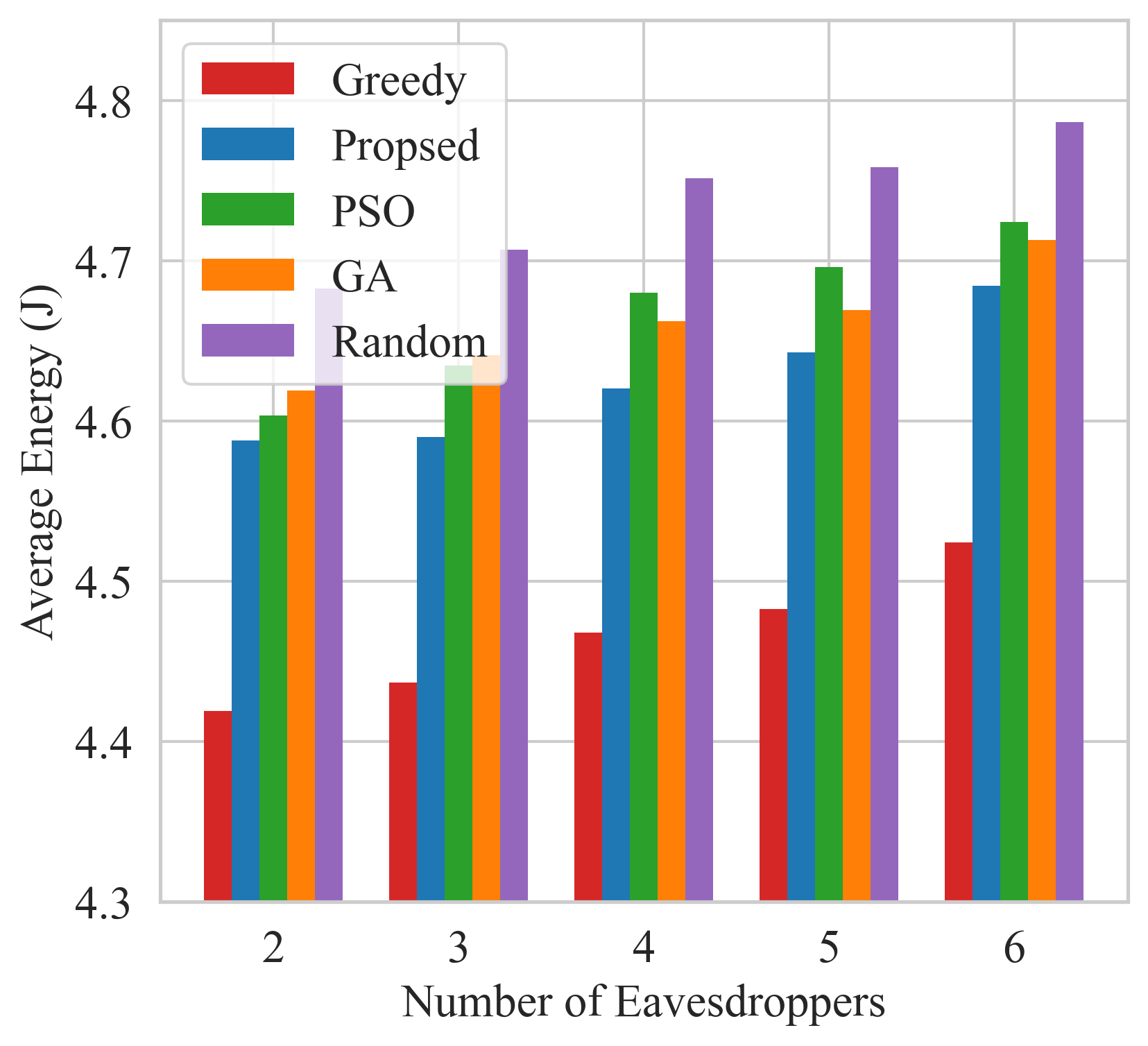}
					\label{Eavesdropper_Energy}
                    \vspace{-5mm} 
			\end{minipage}
	}

	\caption{Performance of different schemes under different number of Eavesdroppers.}
    \label{DRL_Performance}
\end{figure*}
The simulation results presented in Fig.\ref{Converge} demonstrate the effectiveness of the proposed scheme in comparison to PSO and GA algorithms. The proposed scheme exhibits superior convergence performance, achieving a more optimal cost value within approximately 200 iterations. In contrast, the PSO scheme exhibits a slower convergence rate, with the cost value stabilizing after roughly 125 iterations. The GA scheme, while showing some initial improvement, ultimately achieves a less optimal cost value than the proposed scheme. These results highlight the proposed scheme's advantages in terms of convergence speed, stability, and final solution quality, indicating its effectiveness in solving the task scheduling problem.

The radar chart in Fig.\ref{Secure-Legitimate} vividly illustrates the performance of different schemes in terms of reliable transmission rate across varying numbers of legitimate users when eavesdropper number is 5 in the first user association stage. As the number of legitimate user increases, the reliable transmission rate of all four schemes have little change due to the randomness of user position. As can be seen, our proposed scheme achieves the highest reliable transmission rate across all user counts, demonstrating its superiority in enhancing transmission security and reliability. The scheme without AN exhibits lower performance, indicating the critical role of AN in safeguarding transmissions. Due to the without AN scheme can choose LEO satellite with better $R^{\text{sec}}$, it can still achieve a certain level of security. Notably, the random scheme underperforms compared to the greedy scheme. This is likely because the nearest satellite typically offers better SINR for legitimate users, thereby improving the transmission quality and security. When the number of legitimate user is 30, the reliable transmission rate of the proposed scheme is reduced by approximately 77.7\%, 74.5\%, and 50.7\% compared to the random, greedy, and without AN schemes, respectively.

The radar chart in Fig.\ref{Secure-Eavesdropper} provides a clear comparison of the reliable transmission rates achieved by different schemes under varying numbers of eavesdroppers when legitimate number is 30 in the first user association stage. As the number of eavesdroppers increases, the reliable transmission rate of all four schemes shows an downward trend, indicating that a larger number of eavesdroppers leads to higher security risk. It is evident that the proposed scheme consistently outperforms the other schemes across all eavesdropper counts, demonstrating its effectiveness in maintaining a high level of transmission security and reliability even in the presence of multiple eavesdroppers. The performance of without AN scheme highlights the importance of AN in protecting against eavesdropping. The results further reinforce the superiority of the proposed scheme in optimizing user association for secure communications in environments with different numbers of eavesdroppers. When the number of eavesdropper is 12, the reliable transmission rate of the proposed scheme is reduced by approximately 85.2\%, 83.2\%, and 72.4\% compared to the random, greedy, and without AN schemes, respectively.

The simulation results presented in Fig.\ref{Legitimate_Delay} and Fig.\ref{Legitimate_Energy} provide comprehensive insights into the performance of different schemes in terms of average delay and average energy consumption as the number of legitimate users increases when eavesdropper number is 5. In Fig.\ref{Legitimate_Delay}, the delay across schemes shows an upward trend as the number of legitimate users grows, due to more user tasks will result in greater queuing delay and transmission delay. The random scheme exhibits the highest average delay. The greedy scheme also shows a notable increase in delay, though slightly less pronounced than the random scheme. Among all schemes, the proposed scheme demonstrates the lowest average delay. When the number of legitimate user is 25, the average delay of the proposed scheme is reduced by approximately 28.5\%, 18.0\%, 2.2\%, and 4.2\% compared to the random, greedy, GA and PSO schemes, respectively.

Fig.\ref{Legitimate_Energy} illustrates the average energy consumption trends. As the number of legitimate users increases, energy consumption rises across all schemes. The random scheme consistently consumes the most energy. The greedy scheme displays the lowest energy usage, due to all the task are scheduled to the nearest LEO satellite. In contrast, the proposed scheme achieves the second lowest energy consumption, with values consistently below those of other schemes. When the number of legitimate user is 25, the average energy consumption of the proposed scheme is reduced by approximately 5.2\%, 5.0\%, and 4.0\% compared to the random, GA and PSO schemes, respectively.

The simulation results in Fig.\ref{Eavesdropper_Delay} and Fig.\ref{Eavesdropper_Energy} offer valuable insights into how different schemes perform in terms of average delay and average energy consumption as the number of eavesdroppers increases  when legitimate user number is 30. In Fig.\ref{Eavesdropper_Delay}, the average delay shows an little upward trend across all schemes as the number of eavesdroppers grows. This is because as more eavesdroppers emerge, users have fewer options to stay security. The random scheme exhibits the highest average delay. The greedy scheme also shows a notable increase in delay, though it is slightly less pronounced than the random scheme. Among all schemes, the proposed scheme demonstrates the lowest average delay. When the number of eavesdroppers is 6, the average delay of the proposed scheme is reduced by approximately 37.2\%, 22.0\%, 1.2\%, and 1.9\% compared to the random, greedy, GA and PSO schemes, respectively.

Fig.\ref{Eavesdropper_Energy} illustrates the average energy consumption trends. As the number of eavesdroppers increases, energy consumption rises across all schemes. The random scheme consistently consumes the most energy. The greedy scheme still displays the lowest energy usage due to nearest scheduling decision. The proposed scheme achieves the second lowest energy consumption, with values consistently below those of other three schemes. When the number of eavesdroppers is 6, the average energy consumption of the proposed scheme is reduced by approximately 2.6\%, 0.8\%, and 1.1\% compared to the random, GA and PSO schemes, respectively. The delay and energy consumption change little because simply adding eavesdroppers has little impact on user task processing, and the number of users is not increased.

\section{Conclusion}
\label{sec:conclusion}
In this paper, we investigated the secure task scheduling problem in STECN under delay, energy, and physical-layer security considerations. To tackle the challenges posed by dynamic satellite topologies and resource constraints, we proposed a two-stage heuristic framework that jointly optimizes AN power allocation, user association, and task scheduling. The first stage constructs secure communication topologies by maximizing the number of users satisfying the secrecy rate constraint, while the second stage employs a mayfly algorithm to minimize delay and energy consumption under feasible associations. Simulation results demonstrate the effectiveness of our framework.

In future work, we will enhance the proposed framework by integrating adaptive learning-based strategies, such as deep reinforcement learning (DRL) and transformer-based methods, to enable real-time secure scheduling in larger-scale satellite constellations. Additionally, we will explore advanced physical-layer security techniques and multi-layer security mechanisms to further bolster the security of task scheduling in STECNs. We will also focus on energy-efficient computing and battery management to improve the sustainability of the system.

\appendices

\begin{IEEEbiographynophoto}{TING YOU} is currently pursuing the BS degree at Beijing Sport University, Beijing, China. Her current research interests include the Internet of Things, satellite edge computing, and edge caching.
\end{IEEEbiographynophoto}

\begin{IEEEbiographynophoto}{YUHAO ZHENG} received the BS degree in communication engineering from Beijing University of Posts and Telecommunications (BUPT), in 2023. He is currently working toward his master degree in the State Key Laboratory of Networking and Switching
Technology, BUPT, China. His current research interests include satellite-terrestrial integrated networks and satellite edge caching.
\end{IEEEbiographynophoto}

\begin{IEEEbiographynophoto}{KEJIA PENG} is currently pursuing the BS degree at Nanjing University of Science and Technology, China, and Mendeleev University of Chemical Technology of Russia, Moscow, Russia. Her current research interests include industrial Internet of Things and edge caching.
\end{IEEEbiographynophoto}

\begin{IEEEbiographynophoto}{CHANG LIU} received the BS degree in 2013 and the MS degree in 2015 from Peking University, Beijing, China, and the PhD degree from the Technical University of Munich, Munich, Germany, in 2021. He is currently an associate professor with Beijing Sport University, Beijing, China. He has published more than 60 SCI-indexed papers. His research interests include edge computing, information-centric networking, and the industrial Internet of Things.
\end{IEEEbiographynophoto}

\EOD

\end{document}